# Impact of the modulation doping layer on the ν=5/2 anisotropy

X. Shi[1], W. Pan[1], K.W. Baldwin[2], K.W. West[2], L.N. Pfeiffer[2], and D.C. Tsui[2]

[1] Sandia National Labs, Albuquerque, New Mexico 87185, USA
[2] Princeton University, Princeton, New Jersey 08544, USA

Abstract:

We have carried out a systematic study of the tilted magnetic field induced anisotropy at the Landau level filling factor ν=5/2 in a series of high quality GaAs quantum wells, where the setback distance (d) between the modulation doping layer and the GaAs quantum well is varied from 33 to 164 nm. We have observed that in the sample of the smallest d electronic transport is anisotropic when the in-plane magnetic field ($B_{ip}$) is parallel to the [1-10] crystallographic direction, but remains more or less isotropic when $B_{ip}$ // [110]. In contrast, in the sample of largest d, electronic transport is anisotropic in both crystallographic directions. Our results clearly show that the modulation doping layer plays an important role in the tilted field induced ν=5/2 anisotropy.

The modulation doping scheme was invented nearly 40 years ago [1] to achieve high electron mobility in GaAs quantum wells. This invention has had enormous impact on our daily life and scientific discoveries. Indeed, it is hard to imagine any cell phone without high electron mobility transistors, which are the direct outcome of the modulation doping invention.

Scientifically, the introduction and perfection of modulation doping made it possible to increase the electron mobility of the two-dimensional electron system (2DES) from merely ~ 10,000 $cm^2$/Vs in the late 1970's to ~ 40,000,000 $cm^2$/Vs a few years ago [2]. One of many surprising discoveries enabled by this increase in electron mobility is the fractional quantum Hall effect (FQHE) [3], where the 2D electrons form a new incompressible liquid caused by strong electron-electron interactions. Since the first observation of the FQHE at the Landau level filling factor $\nu=1/3$, a total of 80 some FQHE states have been identified and almost all of them can be understood within the Laughlin wave-function [4], hierarchy [5,6], and composite fermion (CF) [7] models.

For a long time, the role of modulation doping on the FQHE was to increase electron mobility and, thus, to magnify the electron-electron interaction effect and to uncover more fragile FQHE states. Recently, however, new roles played by modulation doping on many-body electron phases have been noted. For example, it was proposed that the possible quasiperiodic potential in the modulation doping layer might be the cause of high-frequency magneto-oscillations around $\nu=1/2$ [8]. Moreover, it was observed that long-range charged disorder potential fluctuations originating from modulation doping layers were more detrimental to the stability of the $\nu=5/2$ FQHE state than short-range neutral disorder potential fluctuations [9, 10, 11].

In this paper, we show convincing evidence that modulation doping layers play an important role in the anomalous behavior of in-plane magnetic field ($B_{ip}$) induced anisotropy at the even-denominator 5/2 FQHE state. We report results from a systematic tilted magnetic field study in a series of high quality GaAs quantum wells, in which the impact of modulation doping layers on the 5/2 anisotropic phase is varied by changing the setback distance (d) between the modulation doping layers and the GaAs quantum well. We have observed that in the sample of the shortest d (or strongest impact of modulation doping) electronic transport is anisotropic when $B_{ip}$ is parallel to the [1-10] crystallographic direction, but remains more or less isotropic in the other direction of [110], consistent with previous work [12]. In contrast, in the sample of the largest d (or weakest impact of modulation doping), electronic transport is anisotropic in both crystallographic directions. Our results clearly show that the modulation doping layers do matter in the tilted magnetic field induced anisotropy in the 5/2 FQHE.

Among all the FQHE states, the 5/2 state remains the most exotic. In 1987, a strong minimum in the magnetoresistance $R_{xx}$ and a plateau like feature in the Hall resistance $R_{xy}$ were observed at the even-denominator filling factor $\nu=5/2$ [13]. This state was confirmed unequivocally to be a true FQHE state 12 years later [14]. The observation of this even-denominator FQHE came as a total surprise, as it escapes the so-called odd-denominator rule set by the Laughlin [4], hierarchy [5,6], and composite fermion [7] models. Now, it is generally believed that this state is due to the pairing of CFs [15].

The 5/2 state has been the center of FQHE research for more than 15 years. This state may be a Pfaffian state and, thus, its quasiparticles obey the non-Abelian statistics. As a consequence, the 5/2 state can be used for topological quantum computation (QC) [16], which can have an enormous advantage over other QC approaches where the error rate is relatively large. Extensive experimental work has been carried out to examine the spin polarization of the 5/2 state. The idea is quite straightforward. The 5/2 FQHE ground state must be fully spin polarized if the 5/2 state is a Pfaffian state. One of the most commonly used techniques to examine the spin polarization of a FQHE state is to tilt the sample in the magnetic field [17-22]. If spin polarized, the FQHE state would suffer almost no detrimental effect under tilt. However, if it is spin unpolarized or partially polarized, it can undergo a spin transition. The spin unpolarized FQHE is first destroyed and then reemerges as a spin polarized one. The first tilted magnetic field experiment [17] on the 5/2 state showed that the 5/2 state was quickly weakened and disappeared as the sample was tilted away from the sample normal. This result apparently favored a spin unpolarized ground state at $\nu=5/2$. Later, in two experiments [18,19], it was observed that the disappearance of the 5/2 state was due to the transition from the FQHE state to an anisotropic state, probably a stripe state or a unidirectional charge density wave state. Finite size numerical calculations [23] further showed that the 5/2 FQHE and the anisotropic phases are very close in energy. An added in-plane magnetic field ($B_{ip}$) would make the 2DES effectively thinner and, hence, increase the ratio of $V_1/V_3$, where $V_1$ and $V_3$ are Haldane pseudopotentials [23]. As a result, the anisotropic state becomes stable. Moreover, in these two experiments it was also demonstrated that the orientation of the 5/2 stripes was locked and always perpendicular to the direction of the in-plane magnetic field [18,19,24-26], independent of crystallographic directions. However, in a recent report [12] in an extremely high density sample, this rotational symmetry was broken. Whether or not the tilted magnetic field induced anisotropy exists depended on the GaAs crystallographic directions. Electronic transport became anisotropic when $B_{ip}$ was parallel to [1-10] but remained isotropic if $B_{ip}$ parallel to [110]. This observation is in contrast with the first two experiments, and its physics origin needs further exploration.

To this purpose, we have grown a series of GaAs modulation doped quantum wells with a fixed doping density but a different set-back distance. The well width is kept at 20 nm, in order that only the lowest subband is occupied. The schematic of their growth structures is shown in Fig. 1(a). This layer structure is the same for all the samples except for the set-back distance, d, which is varied from 33 to 164 nm. In Fig. 1(b) we show the electron density and mobility as a function of the setback distance. It can be seen that the electron density increases monotonically with decreasing d, while the mobility shows a non-monotonic d dependence, reaching a maximal value around d = 52 nm. In Fig. 1(c), we show $R_{xx}$ and $R_{yy}$ in the d=52nm sample at zero tilt. Strong anisotropy is observed at half fillings $\nu=9/2$ and 11/2 in the higher Landau levels. The hard (high resistance) axis is along [1-10] and the easy (low resistance) axis along [110]. In contrast, electron transport is isotropic at $\nu=5/2$ and 7/2 in the second Landau level. Finally, no anisotropy is observed in the Hall resistance. These results are consistent with previous work [27-36].

In Fig. 2, we show the tilted magnetic field dependence of $R_{xx}$ and $R_{yy}$ measured in the d=164nm sample in two configurations. In the first configuration, the in-plane magnetic field $B_{ip}$ is parallel to [1-10] (or perpendicular to [110]). It can be seen in Fig. 2(a) that $R_{xx}$ and $R_{yy}$ are almost the same in the regime of $4 > \nu > 2$ at zero tilt angle. With increasing tilt angle or $B_{ip}$, the $R_{xx}$ minimum at $\nu=5/2$, measured along $B_{ip}$, increases and the QHE feature becomes weakened. By 75.9$^o$, a giant peak has developed at $\nu=5/2$ in $R_{xx}$. The $R_{yy}$ at $\nu=5/2$, measured perpendicular to $B_{ip}$, remains a dip and displays a very weak $B_{ip}$ dependence. In Fig. 2(b), we show $R_{xx}$ and $R_{yy}$ at $\nu=5/2$ as a function of $B_{ip}$. Again, $R_{xx} \sim R_{yy}$ in the perpendicular B field (or $B_{ip}=0$). Under in-plane magnetic fields, $R_{xx}$ and $R_{yy}$ become anisotropic.

We then pulled out the sample and rotated it by 90$^o$ and re-cooled it using the same cooling procedure. In this configuration [Figs. 2(c) and 2(d)], the in-plane field is parallel to [110] (or perpendicular to [1-10]). Again, $R_{xx}$ and $R_{yy}$ are virtually the same at the zero tilt. Upon tilting, anisotropy in $R_{xx}$ and $R_{yy}$ develops with increasing tilt angle, the same as in the first configuration.

We summarize here that in the d=164 nm quantum well the 5/2 state becomes anisotropic in both configurations, and the hard axis is always parallel to $B_{ip}$, independent of crystallographic directions. These results are consistent with those reported in Refs. [18,19].

We then examined two more samples of different d. In Fig. 3, we show the results in the sample with the smallest d of 33 nm. When $B_{ip}$ is parallel to the [1-10] direction [Figs. 3(a) and 3(b)], $R_{xx}$ and $R_{yy}$ display a similar in-plane field induced anisotropy as in Figs. 2 (a) and s(b). When $B_{ip}$ is applied in the [110] direction [Figs. 3(c) and 3(d)], the 5/2 state is also destroyed. However, it remains isotropic even under high in-plane fields. Indeed, $R_{xx} \sim R_{yy}$, even when $B_{ip}$ is larger than 8 T. This is very different from that in Figs. 2 (c) and 2(d).

To summarize the results in Fig. 3, we observe that in the d=33nm sample the 5/2 FQHE state is destroyed by in–plane magnetic fields. It becomes anisotropic when $B_{ip}$ is parallel to [1-10], but remains isotopic when $B_{ip}$ is in the [110] direction. This observation is consistent with the findings reported in Ref. [12].

In Fig. 4, we plot the anisotropy factor (AF), defined as AF = $(R_{xx}-R_{yy})/(R_{xx}+R_{yy})$, as a function of decreasing d or increasing impact of modulation doping. The anisotropy develops in both in-plane field directions for d = 164 nm. As d is reduced to 52 nm, the anisotropy is fully developed (or AF $\approx$ 1) when $B_{ip}$ is in the [1-10] direction. However, AF saturates to a value of around 0.5 when $B_{ip}$ // [110]. With d further reduced to 33 nm, AF is now almost zero for $B_{ip}$ // [110] while it is close to 1 for $B_{ip}$//[1-10].

Our results in Fig. 4 clearly demonstrate the distance between the modulation doping layers and quantum well is a determining factor in causing the anomalous $B_{ip}$ induced anisotropy in high density samples. In the following we consider two previously proposed mechanisms [8, 37-39] that can explain this anomaly in the samples where the modulation doping layer effect is strong.

First, it was argued in Ref. [38] that the electric field between the 2DES and the modulation doping layers [37-39] could generate an anisotropic band mass, and this anisotropic mass can provide a symmetry breaking mechanism in the magnetic field. As shown in Ref. [38], due to the "uniaxial stress" induced by this electric field, the GaAs bonds in the [110] direction are stretched, while the bonds in the [1-10] direction are shortened. As a consequence, the effective band mass is heavier in the [1-10] direction and lighter in [110], which favors the stripes parallel to [110]. It can be expected that when d is large or n is low, the pinning force due to the band mass anisotropy is weaker than the de-pinning force induced by $B_{ip}$. Consequently, the stripes are locked perpendicularly to the direction of $B_{ip}$, as observed in the past. However, when d becomes smaller or n higher, the band mass anisotropy and, accordingly, the [110] pinning force increase. When $B_{ip}$ is along [1-10], both $B_{ip}$ and mass anisotropy help align the stripes perpendicular to $B_{ip}$. When $B_{ip}$ is pointed to the [110] direction, the re-orientation of stripes to [1-10] due to $B_{ip}$ competes with the pinning to [110] due to the band mass anisotropy, which is now stronger due to higher n. As a result, it is possible that the stripes finally align themselves at an angle between the [110] and [1-10] directions. This can give rise to an apparent isotropic electron transport.

For the second mechanism, we consider a possible quasiperiodic potential in the modulation doping layers by electron correlation [8]. Surface morphology studies [40-42] suggested that the periodic potential lines are in the [1-10] direction. It is known that this periodic potential can cause an effect similar to an artificially modulated sample [8,40], and helps orientate the stripes perpendicular to the potential modulation [40,41,43,44]. When d is large, this periodic potential is weak and the reorientation of stripes in the presence of $B_{ip}$ is dominant. As a consequence, the in-plane field induced stripes are determined by the direction of $B_{ip}$. When d is small, however, the periodic potential can provide a much stronger pinning force. With $B_{ip}$ in the [1-10] direction, or parallel to the potential modulation lines, pinning due to both $B_{ip}$ and this periodic potential helps pin the stripes perpendicular to $B_{ip}$. When $B_{ip}$ is perpendicular to the potential modulation or in the [110] direction, the orientation of the 5/2 stripes is now determined by the competition between $B_{ip}$ depinning and the quasiperiodic potential pinning. The outcome of this competition may again align the stripes along a direction between [110] and [1-10], giving rise to an isotropic electron transport.

In summary, we report results from a systematic tilted magnetic field study of the ν=5/2 FQHE in a series of high quality GaAs quantum wells, in which the setback distance between the modulation doping layers and GaAs quantum well is varied from 33 to 164nm. We have observed that in the sample of the shortest d, electronic transport is anisotropic when $B_{ip}$ is parallel to the [1-10] crystallographic direction but remains more or less isotropic in the other direction of [110], consistent with previous work. In contrast, in the samples of larger d, electronic transport is anisotropic in both crystallographic directions. Our results clearly show that the modulation doping layers do matter in the tilted magnetic field induced anisotropy in the 5/2 FQHE.

We thank Michael Manfra for helpful discussions. This work was supported by the Department of Energy, Office of Basic Energy Sciences, Division of Materials Sciences and Engineering. Sandia National Laboratories is a multiprogram laboratory managed and operated by Sandia Corporation, a wholly owned subsidiary of Lockheed Martin Corporation, for the U.S. Department of Energy's National Nuclear Security Administration under Contract No. DE-AC04-94AL85000. Sample growth at Princeton University was partially funded by the Gordon and Betty Moore Foundation, Keck Foundation, as well as the National Science Foundation MRSEC Program through the Princeton Center for Complex Materials. A portion of the work was performed at the National High Magnetic Field Laboratory, which is supported by the National Science Foundation (DMR-1157490), the State of Florida, and the DOE. We thank Tim Murphy, Ju-Hyun Park, and Glover Jones for experimental help.

Figure captions:

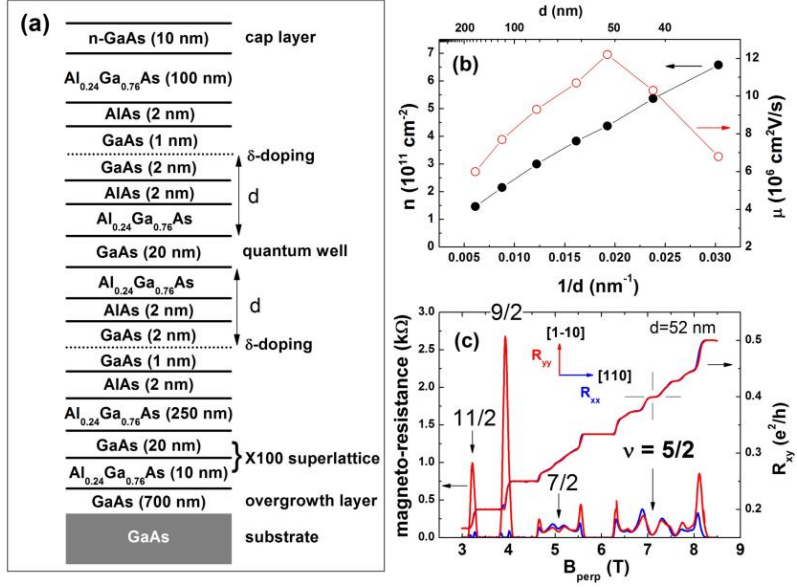

FIG.1. (Color online) (a) Schematic of GaAs quantum well growth structure. The setback distance (d) is defined between the edge of the GaAs quantum well and the δ-doping layer. (b) Electron density (n) and mobility (μ) as a function of 1/d. We plot in (c) the $R_{xx}$, $R_{yy}$, and $R_{xy}$ traces in the d=52 nm sample, where the electron mobility is the highest.

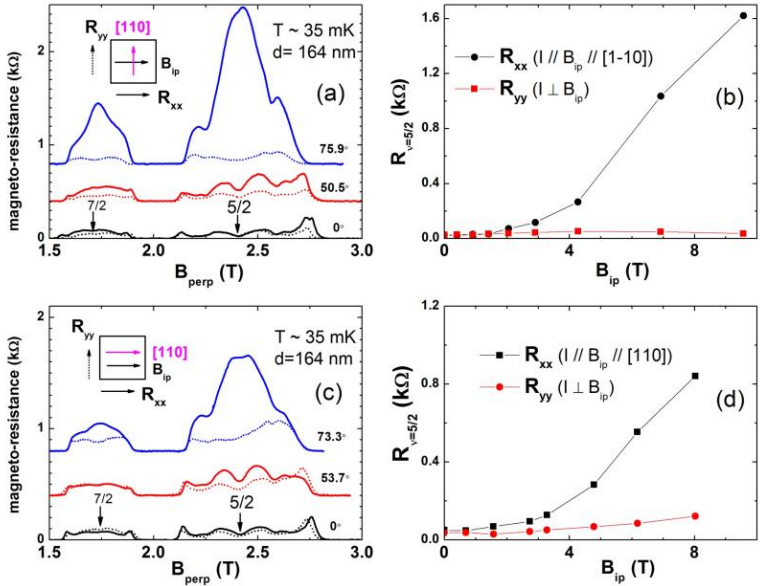

FIG.2. (Color online) (a) $R_{xx}$ and $R_{yy}$ measured at various tilt angles in the d=164nm sample. The arrows mark the 5/2 and 7/2 states. The in-plane magnetic field ($B_{ip}$) direction is parallel to the [1-10] direction. In (b) we plot the $R_{xx}$ and $R_{yy}$ values at ν=5/2 as a function of $B_{ip}$. (c) is similar to (a), but for $B_{ip}$ // [110]. (d) is similar to (b), for $B_{ip}$//[110].

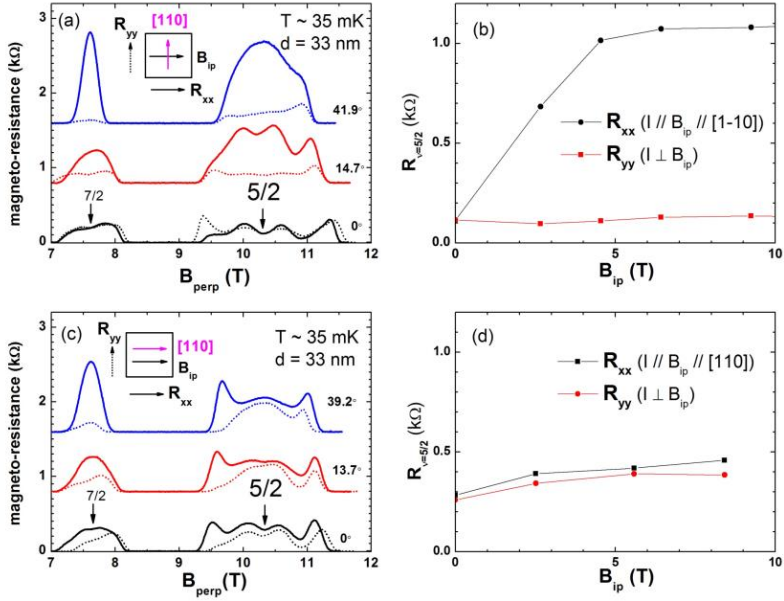

FIG.3. (Color online) (a) $R_{xx}$ and $R_{yy}$ measured at various tilt angles in the d=33nm sample. The arrows mark the 5/2 and 7/2 states. The in-plane magnetic field ($B_{ip}$) direction is parallel to the [1-10] direction. In (b) we plot the $R_{xx}$ and $R_{yy}$ values at ν=5/2 as a function of $B_{ip}$. (c) is similar to (a), but for $B_{ip}$ // [110]. (d) is similar to (b), for $B_{ip}$/[110].

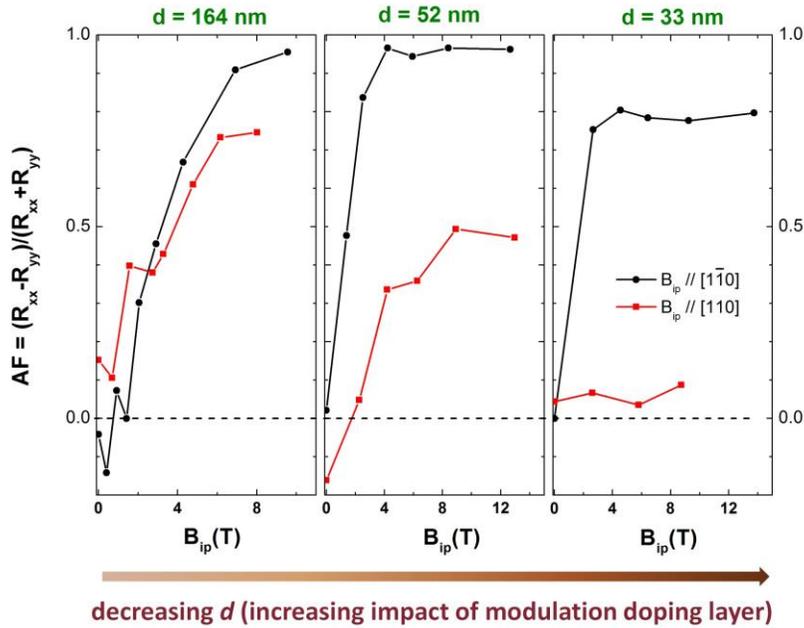

FIG.4. (Color online) The anisotropy factor AF at ν=5/2, defined as AF = $(R_{xx}-R_{yy})/(R_{xx}+R_{yy})$, for three selected samples, where the impact of the modulation doping layer increases with decreasing d.